\documentclass[sigconf]{acmart}
\AtBeginDocument{%
  \providecommand\BibTeX{{%
    \normalfont B\kern-0.5em{\scshape i\kern-0.25em b}\kern-0.8em\TeX}}}

\setcopyright{rightsretained}
\copyrightyear{2024}
\acmYear{2024}
\acmConference{SIGGRAPH Posters '24}{July 27 - August 01, 2024}{Denver, CO, USA}\acmBooktitle{Special Interest Group on Computer Graphics and Interactive Techniques Conference Posters (SIGGRAPH Posters '24), July 27 - August 01, 2024}\acmDOI{10.1145/3641234.3671036}
\acmISBN{979-8-4007-0516-8/24/07}

\setcopyright{rightsretained}
\copyrightyear{2024}
\acmYear{2024}
\acmConference{SIGGRAPH Posters '24}{July 27 - August 01, 2024}{Denver, CO, USA}\acmBooktitle{Special Interest Group on Computer Graphics and Interactive Techniques Conference Posters (SIGGRAPH Posters '24), July 27 - August 01, 2024}\acmDOI{10.1145/3641234.3671036}
\acmISBN{979-8-4007-0516-8/24/07}

\begin{document}

\title{Projecting Radiance Fields to Mesh Surfaces}

\author{Adrian Xuan Wei Lim}
\affiliation{%
  \institution{Roblox}
   \country{USA}
}
\email{xlim@roblox.com}

\author{Lynnette Hui Xian Ng}
\affiliation{%
  \institution{Carnegie Mellon University}
   \country{USA}
}
\email{lynnetteng@cmu.edu}

\author{Nicholas Kyger}
\affiliation{%
  \institution{Roblox}
   \country{USA}
}
\email{nkyger@roblox.com}

\author{Tomo Michigami}
\affiliation{%
  \institution{Roblox}
   \country{USA}
}
\email{tmichigami@roblox.com}

\author{Faraz Baghernezhad}
\affiliation{%
  \institution{Roblox}
   \country{USA}
}
\email{fbaghernezhad@roblox.com}

\renewcommand{\shortauthors}{Lim et al.}

\begin{abstract}
Radiance fields produce high fidelity images with high rendering speed, but are difficult to manipulate. We effectively perform avatar texture transfer across different appearances by combining benefits from radiance fields and mesh surfaces. We represent the source as a radiance field using 3D Gaussian Splatter, then project the Gaussians on the target mesh. Our pipeline consists of Source Preconditioning, Target Vectorization and Texture Projection. The projection completes in 1.12s in a pure CPU compute, compared to baselines techniques of Per Face Texture Projection and Ray Casting (31s, 4.1min). This method lowers the computational requirements, which makes it applicable to a broader range of devices from low-end mobiles to high end computers.
\end{abstract}



\begin{teaserfigure}
  \includegraphics[width=\textwidth]{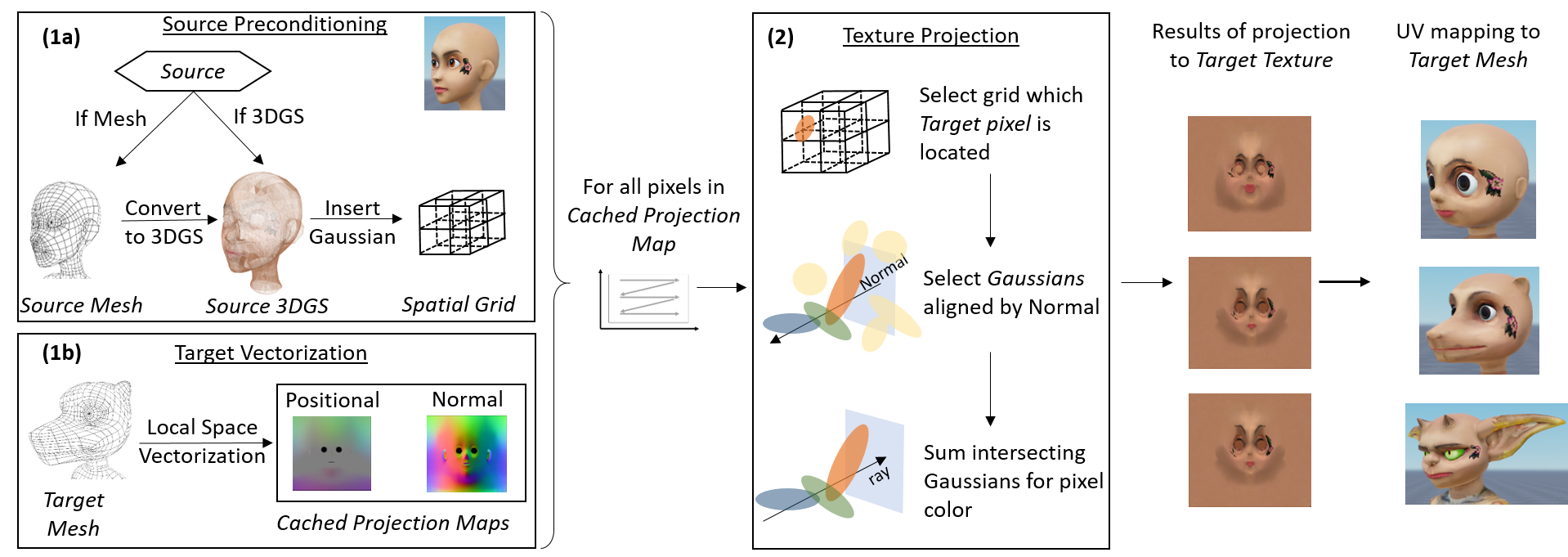}
  \caption{Projection Pipeline, with example of texture transfer from one source avatar to three differently shaped target meshes.}
  \label{fig:Framework}
\end{teaserfigure}


\maketitle

\section{Introduction}
Texture transfer between differently shaped avatars is an important rendering problem. Work on texture transfer includes mesh methods such as using adaptive triangulations of meshes, which completes 11 texture transfers in 38 minutes \cite{schmidt2023surface}; or mapping pixels from the UV parameterization of a surface to the 3D avatar which was used to transfer clothing onto avatars \cite{mir2020learning}; or deep learning methods that perform direct image-to-image translation \cite{shysheya2019textured}.
However, these methods take a long time to process and render the textures.

To speed up the texture transfer process, radiance field rendering methods using 3D Gaussians (3DG) can be harnessed. 3DG  can be used to preserve the continuous properties of the original texture, and avoid unnecessary computation in empty spaces \cite{kerbl20233d}. Most importantly, such a technique preserves rendering quality.
\citet{kerbl20233d} used 3DG as volumetric representation for real-time scene rendering.
\citet{wang20243d} generated 3DG Textures for volumetric texturing in animations. 

In this work, we present a method for avatar texture transfer. The texture transfer is mediated through a Gaussian splattering-based pipeline. 
We represent our source radiance field as a 3D Gaussian Splatter, and incorporated local space projection \citet{10.1145/3588028.3603653} to cache and speed up locality projection. We then directly project the Gaussians on the target texture, blending them for colour and lighting. Our technique efficiently completes realistic avatar texture transfer across different appearances in 1.12s.

\section{Methodology}
\autoref{fig:Framework} shows the Projection Pipeline and the resultant outputs for three target shapes from a single source mesh. The figure illustrates transferring face tattoo and makeups from the face of one avatar to three different meshes of different shapes.

\paragraph{Source Preconditioning.} We accept two source types: (1) 3D polygon mesh, and (2) a 3DGS. To convert a 3D mesh into 3DGS, we generated a Gaussian per pixel of its texture. The Gaussian's position and normal corresponds to the current pixel local space. The covariance spread is determined by its rate of change of its positional iteration. The colour corresponds to the current pixel colour. We then densify the 3DGS to have uniform surface density and to increase the overlap between Gaussians. 

We construct a Spatial Grid to contain $\sim$1 target pixel per subgrid. The grid size is cube root of total target pixel count. 
The Gaussians are added into all subgrids of the Spatial Grid that collides with their covariance volume. This grid setup converts a polynomial to constant search at projection time by caching pixel locality information during creation, eliminating the need for a global search for neighbouring Gaussians for each pixel during projection.

\paragraph{Target Vectorization.} We rasterize all triangles in the target mesh to their UV-space, converting the mesh into Cached Projection Maps. Those texture maps contains the local space position and normal for all target pixels. There is a one-to-one size match between our Cached Projection Maps and our target texture to allow for constant space and time lookups. 

\paragraph{Texture Projection.} For each pixel in the Cached Projection Map, we index the subgrid based on its position. Next, we filter for Gaussians in the indexed subgrid that are aligned with the pixel with a dot product between each Gaussian and the normal. We then perform a ray cast from the target position and normal. The colours of the Gaussians intersecting the ray are blended by their alpha values, and are weighted by their angle to the normal, forming the colour of the target texture pixel.

\section{Results and Discussion}

\paragraph{Performance.} Our experiments used a single-thread CPU (Intel i9-12900H) implementation. Given a mesh with $\sim$10k triangles, $\sim$1m Gaussians and a texture size of 1024x1024, our pipeline completed in $1.12\pm0.05$s. This timing result was recorded from differing source and targets with similar specifications. 

The pipeline can be sped up further by parallelizing the following steps: Spatial Grid insertion, Local Space Vectorization, and the Target Projection.
Experimenting with thread counts of 2 and 4, the pipeline completes in 0.68s and 0.46s respectively. Therefore, using parallelizion, we achieve near linear decrease in processing time.

Our pipeline is much faster compared to baseline methods on the same CPU: 31sec for Per Face Texture Projection and 4.1min for Ray Cast technique. Rendering quality is not sacrificed in the speed gain, illustrated by the examples in \autoref{fig:Framework} which presents high fidelity images.

\paragraph{Accuracy.} We ran a test on 5 different mesh-based avatar heads. We projected these avatar heads onto themselves to extract the transferred texture. We then compared the result of the transferred texture with the original texture. The two textures are $\sim98\%$ similar. The errors could be introduced during inaccuracies propagated in the 3DGS densification step. 

\paragraph{Deployability.} Our test GPU implementation achieved real-time performance but low-end mobile GPUs do not support rasterization to floating point textures, thus unable to compute Cached Projection Maps. However, a pure CPU approach allows deployment to low-end devices, opening possibilities to leverage on existing compute cloud infrastructure.


\paragraph{Limitations.} Texture transfer to targets that have different mesh shapes from the source will result in a distortion of projection. This is similar to artifacts generated during a planar projection on a sphere. This can be resolved by enclosing the target in a cage and applying a warp deformer to closely match the shape of the target to the source. 

In a mirrored or tiled target, texture transfer with our pipeline can be duplicated or overridden depending on the order of execution of the transfer. A possible resolution to this is to UV-unwrap the target mesh to be UV-space unique. 

\paragraph{Future Work.} Future work involves object masking for the removal of facial accessories, so that only textures are mapped over, and content-aware fill for predicting eye colors during projection. These enhancements will create more photorealistic avatars.

\paragraph{Conclusions} Our projection pipeline efficiently transfers textures across source-texture pairs of different shapes by combining the high fidelity benefits of radiance fields with the manipulability of polygon meshes, opening possibilities to animate the target model. 


\begin{acks}
Much thanks to Charlie Kubal, Jovanni Cutigni, Morgan McGuire, Victor Zordan, and the amazing folks at Roblox and CMU for their support in this research.
Special thanks to Roblox's Game Engine Group, Avatar Personalization Team, Rendering team for support in implementation.
\end{acks}

\bibliographystyle{ACM-Reference-Format}
\bibliography{main}


\begin{thebibliography}{6}


\ifx \showCODEN    \undefined \def \showCODEN     #1{\unskip}     \fi
\ifx \showDOI      \undefined \def \showDOI       #1{#1}\fi
\ifx \showISBNx    \undefined \def \showISBNx     #1{\unskip}     \fi
\ifx \showISBNxiii \undefined \def \showISBNxiii  #1{\unskip}     \fi
\ifx \showISSN     \undefined \def \showISSN      #1{\unskip}     \fi
\ifx \showLCCN     \undefined \def \showLCCN      #1{\unskip}     \fi
\ifx \shownote     \undefined \def \shownote      #1{#1}          \fi
\ifx \showarticletitle \undefined \def \showarticletitle #1{#1}   \fi
\ifx \showURL      \undefined \def \showURL       {\relax}        \fi
\providecommand\bibfield[2]{#2}
\providecommand\bibinfo[2]{#2}
\providecommand\natexlab[1]{#1}
\providecommand\showeprint[2][]{arXiv:#2}

\bibitem[Kerbl et~al\mbox{.}(2023)]%
        {kerbl20233d}
\bibfield{author}{\bibinfo{person}{Bernhard Kerbl}, \bibinfo{person}{Georgios Kopanas}, \bibinfo{person}{Thomas Leimkühler}, {and} \bibinfo{person}{George Drettakis}.} \bibinfo{year}{2023}\natexlab{}.
\newblock \bibinfo{title}{3D Gaussian Splatting for Real-Time Radiance Field Rendering}.
\newblock
\newblock
\showeprint[arxiv]{2308.04079}~[cs.GR]


\bibitem[Lim et~al\mbox{.}(2023)]%
        {10.1145/3588028.3603653}
\bibfield{author}{\bibinfo{person}{Adrian Xuan~Wei Lim}, \bibinfo{person}{Lynnette Hui~Xian Ng}, \bibinfo{person}{Conor Griffin}, \bibinfo{person}{Nicholas Kryer}, {and} \bibinfo{person}{Faraz Baghernezhad}.} \bibinfo{year}{2023}\natexlab{}.
\newblock \showarticletitle{Reverse Projection: Real-Time Local Space Texture Mapping}. In \bibinfo{booktitle}{\emph{ACM SIGGRAPH 2023 Posters}} (Los Angeles, CA, USA) \emph{(\bibinfo{series}{SIGGRAPH '23})}. \bibinfo{publisher}{Association for Computing Machinery}, \bibinfo{address}{New York, NY, USA}, Article \bibinfo{articleno}{48}, \bibinfo{numpages}{2}~pages.
\newblock
\showISBNx{9798400701528}
\urldef\tempurl%
\url{https://doi.org/10.1145/3588028.3603653}
\showDOI{\tempurl}


\bibitem[Mir et~al\mbox{.}(2020)]%
        {mir2020learning}
\bibfield{author}{\bibinfo{person}{Aymen Mir}, \bibinfo{person}{Thiemo Alldieck}, {and} \bibinfo{person}{Gerard Pons-Moll}.} \bibinfo{year}{2020}\natexlab{}.
\newblock \showarticletitle{Learning to transfer texture from clothing images to 3d humans}. In \bibinfo{booktitle}{\emph{Proceedings of the IEEE/CVF Conference on Computer Vision and Pattern Recognition}}. \bibinfo{pages}{7023--7034}.
\newblock


\bibitem[Schmidt et~al\mbox{.}(2023)]%
        {schmidt2023surface}
\bibfield{author}{\bibinfo{person}{Patrick Schmidt}, \bibinfo{person}{D\"orte Pieper}, {and} \bibinfo{person}{Leif Kobbelt}.} \bibinfo{year}{2023}\natexlab{}.
\newblock \showarticletitle{Surface Maps via Adaptive Triangulations}.
\newblock \bibinfo{journal}{\emph{Computer Graphics Forum}} \bibinfo{volume}{42}, \bibinfo{number}{2} (\bibinfo{year}{2023}).
\newblock


\bibitem[Shysheya et~al\mbox{.}(2019)]%
        {shysheya2019textured}
\bibfield{author}{\bibinfo{person}{Aliaksandra Shysheya}, \bibinfo{person}{Egor Zakharov}, \bibinfo{person}{Kara-Ali Aliev}, \bibinfo{person}{Renat Bashirov}, \bibinfo{person}{Egor Burkov}, \bibinfo{person}{Karim Iskakov}, \bibinfo{person}{Aleksei Ivakhnenko}, \bibinfo{person}{Yury Malkov}, \bibinfo{person}{Igor Pasechnik}, \bibinfo{person}{Dmitry Ulyanov}, {et~al\mbox{.}}} \bibinfo{year}{2019}\natexlab{}.
\newblock \showarticletitle{Textured neural avatars}. In \bibinfo{booktitle}{\emph{Proceedings of the IEEE/CVF Conference on Computer Vision and Pattern Recognition}}. \bibinfo{pages}{2387--2397}.
\newblock


\bibitem[Wang and Sin(2024)]%
        {wang20243d}
\bibfield{author}{\bibinfo{person}{Xiangzhi~Eric Wang} {and} \bibinfo{person}{Zackary P.~T. Sin}.} \bibinfo{year}{2024}\natexlab{}.
\newblock \bibinfo{title}{3D Gaussian Model for Animation and Texturing}.
\newblock
\newblock
\showeprint[arxiv]{2402.19441}~[cs.GR]


\end{thebibliography}

\end{document}